\documentclass[10pt,letterpaper]{article}
\usepackage[top=0.85in,footskip=0.75in,marginparwidth=2in]{geometry}
\usepackage[justification=centering]{caption}

\usepackage[utf8]{inputenc}

\usepackage{cite}

\usepackage{nameref,hyperref}

\usepackage{microtype}
\DisableLigatures[f]{encoding = *, family = * }

\usepackage{indentfirst}

\usepackage{changepage}

\usepackage[aboveskip=1pt,labelfont=bf,labelsep=period,singlelinecheck=off]{caption}

\usepackage{lastpage,fancyhdr,graphicx}
\usepackage{epstopdf}
\fancyheadoffset[L]{2.25in}
\fancyfootoffset[L]{2.25in}

\usepackage{color}

\definecolor{Gray}{gray}{.25}

\usepackage{graphicx}

\begin{document}
\vspace*{0.35in}

\begin{flushleft}
{\Large
\textbf\newline{\bf DroidMeter: A Measurement Tool to Study Embedded Web Pages}
}
\newline
\\
Deyu Tian\textsuperscript{1}, tiandeyu@pku.edu.cn\\
\bigskip
\bf{1} Key Lab of High-Confidence Software Technology, MoE (Peking University), Beijing, China\\
\bigskip

\end{flushleft}

\section*{Abstract}

Traditional Web browsing involves typing a URL on a browser and loading a Web page.
In contrast, there is another form of Web browsing on mobile devices, i.e., embedded Web browsing, which occurs when mobile apps embed a Web page within the app.
When the user navigates to the specific page in the app, the Web page is loaded from within the app.
However, little is known about the prevalence or performance of these embedded Web pages.
To analyze the embedded Web browsing performance at scale, we design and implement DroidMeter, a tool that can automatically search for embedded Web pages inside apps, trigger page loads, and retrieve performance metrics.


\section{Introduction}\label{sec:introduction}


Native mobile applications (a.k.a., apps) and mobile Web pages are two of the most popular ways for users to access mobile content~\cite{Serrano:Software2013,Ma:TMC2018}.
While these two medium of access may seem distinct, it is becoming increasingly common for mobile apps to embed a Web page within them.

\begin{figure}
    \centering
    \includegraphics[width=0.18\textwidth]{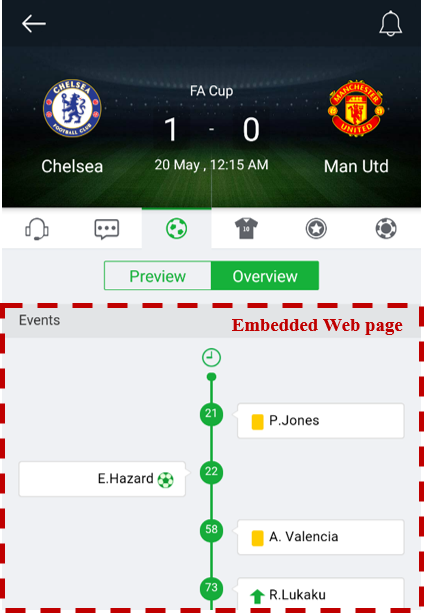}
    \caption{An example of embedded Web page. The part in the dashed line box is an embedded page.}~\label{fig:example}
    \vspace{-2em}
\end{figure}

We call Web pages embedded in mobile apps, {\em embedded Web pages}, in the rest of this paper.
When users navigate to the embedded Web page inside an app, they can interact with the page as they usually do on their browser.
Figure~\ref{fig:example} shows an example of embedded pages for the app {\em All Football}.
Interestingly, part of the screen (indicated by the red dashed lines) is a Web page loaded in the embedded browser, while the rest of the screen is rendered using the native app.
Mobile apps embed Web pages for various reasons including portability across platforms, reusing third-party content easily, and for uniform interface for the users.

Unfortunately, characterizing the performance of the embedded Web page loads is inherently not straightforward but quite challenging.
Traditional Web page load performance can be studied at scale by loading the Web page URL on the browsers programmatically.
However, loading an embedded Web page requires that we navigate to the application page that contains the embedded Web page.
Triggering this embedded page load involves navigating through multiple UI screens on the app, requiring user inputs that are not known {\em a-priori}.
Further, embedded browsers do not support APIs for developer tools that are available in most modern browsers, making performance analysis more challenging.

To address the issue, we design and implement \textit{DroidMeter}, an automation tool of measuring the performance of embedded Web browsing for Android apps.
DroidMeter first explores the app by clicking on UI elements to navigate to various pages, searching for embedded Web pages.
After exploration, for each visited embedded Web page, DroidMeter generates a replay script that contains the sequence of UI events to trigger the embedded Web page.
The key idea of DroidMeter is that we use the structure of UI tree to identify different app pages and thus every single event can be correctly triggered even when there are trivial UI changes during the replay phase.
As for retrieving embedded page's information, we port existing developer tools that are available for traditional browsers to DroidMeter.
Specifically, we re-implement the Chrome remote debugging~\cite{ChromeRemote} and expose the APIs.
This porting allows us to collect performance metrics about the embedded page load, including the load time and the information about Web resources that are loaded.

The remainder of this paper is organized as follows.
Section~\ref{sec:background} shows some background knowledge of how Web pages are embedded on Android.
Section~\ref{sec:methodology} describes the DroidMeter tool that we design for measuring embedded Web browsing.
Section~\ref{sec:casestudy} shows the usage and behavior of DroidMeter by examining its execution procedure.
Section~\ref{sec:related} surveys related work, and Section~\ref{sec:conclusion} concludes the paper.

\section{Background}\label{sec:background}



\begin{figure}
    \centering
    \includegraphics[width=0.6\textwidth]{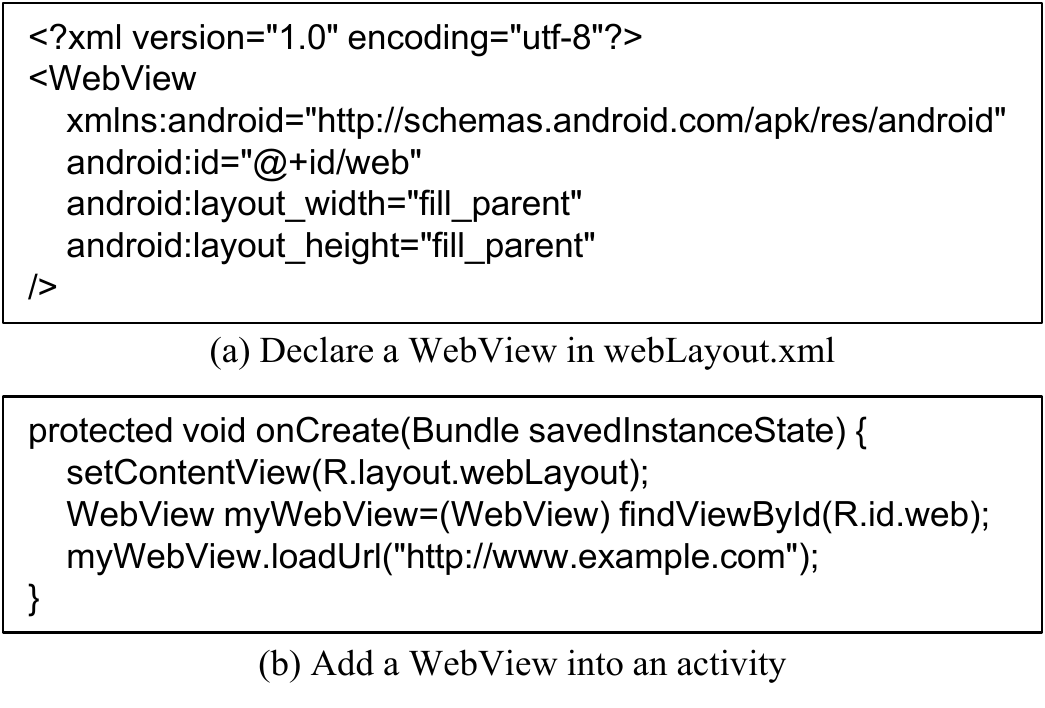}
    \caption{An example of implementing WebView in Android apps.}~\label{fig:codeexample}
    \vspace{-2em}
\end{figure}


Modern mobile OSes such as Android and iOS provide a special UI framework to display Web pages from within an app. For example, Android apps embed, WebView~\cite{WebView}, a framework to load Web pages. Using this framework, one can display Web pages from within the apps using either the Chromium browser engine (after Android 4.4) or the WebKit  engine (before Android 4.4). Figure~\ref{fig:example} shows an example of an app page with WebView. The part highlighted by a dashed line box is an embedded Web page, showing the events of a football match.

Android Webview has been a critical component in Android System for long. The Android Webview is supported by WebKit~\cite{webkit} engine until Android 4.3. Starting from Android 4.4 Kitkat, the Android Webview uses the chromium~\cite{chromium} engine, which is similar to Google Chrome. After Android 5.0 Lollipop, Android Webview becomes a separate app and gets its update via App Store like normal apps. Such modification is aimed to fix some security vulnerabilities noticed in older version.

Figure~\ref{fig:codeexample} shows how an Android app embeds a WebView in an app page. Each page in an Android app is an instance of activity and there is one or more layout files corresponding to the activity. To embed a Webpage, developers first declare a WebView in the layout file. As shown in Figure~\ref{fig:codeexample} (a), a WebView with the id ``\texttt{web}'' is declared in \texttt{webLayout.xml} file. Then in the \texttt{onCreate()} method of the activity, the developer retrieves the WebView instance and loads the page \url{http://www.example.com}. 

\section{Methodology}\label{sec:methodology}



Given the prevalence of embedded Web pages, it is important to study the performance and characteristics of embedded Web pages. There are several developer tools available to measure the performance of traditional browsers. Unfortunately, measuring the performance of embedded Web pages is more challenging.

 First, embedded Web browsing is triggered only if the user reaches the specific activity in the app through a series of UI interactions. These UI interactions are different for different apps, and cannot be obtained {\em a priori}. Second, the developer tools designed to analyze Web performance cannot be directly used to analyze embedded browsing. Although the Chrome browser provides a GUI-based tool to remotely debug embedded pages, they cannot be accessed programmatically.

\subsection{DroidMeter Architecture}

\begin{figure}
    \centering
    \includegraphics[width=0.8\textwidth]{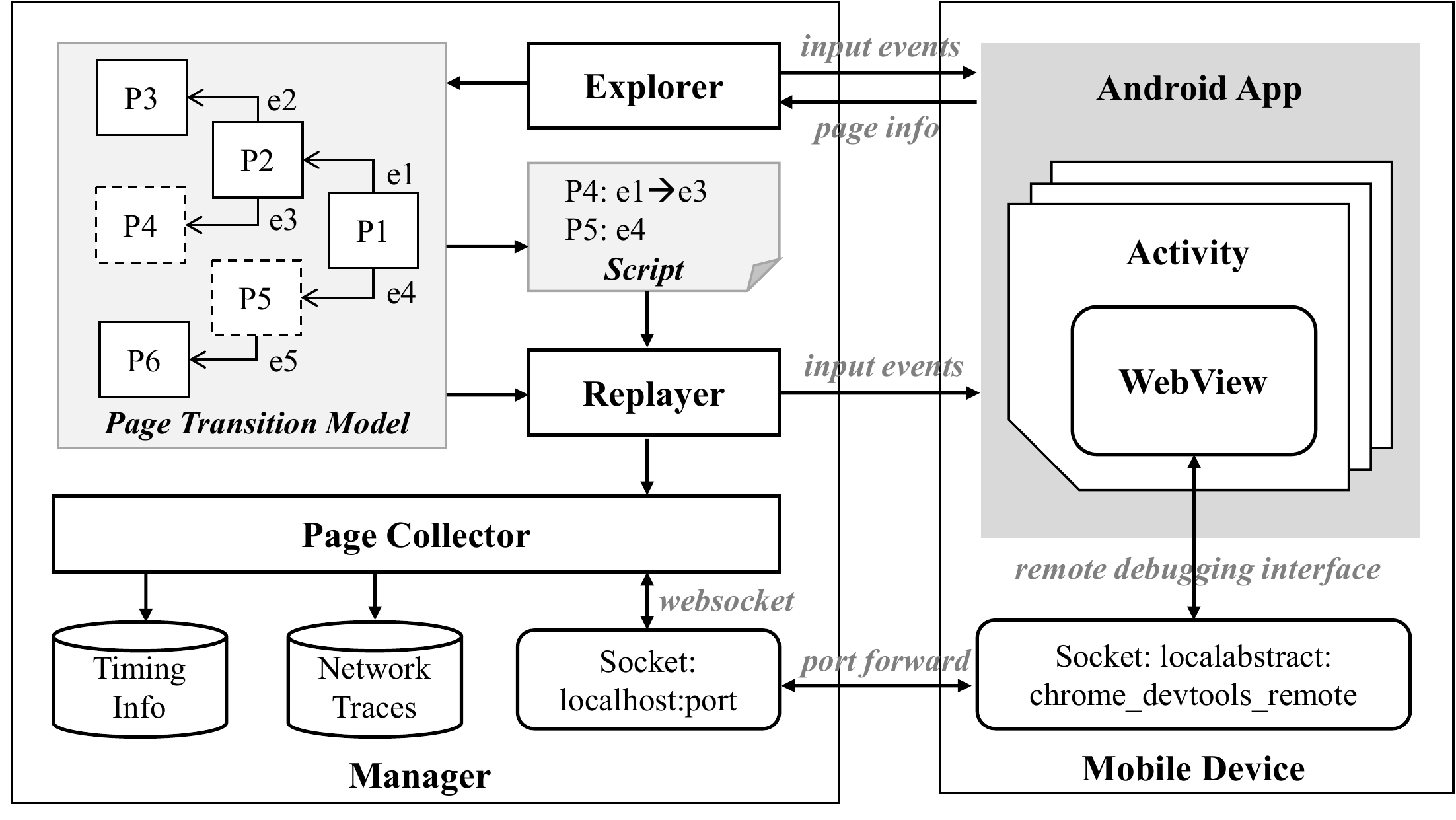}
    \caption{Architecture of DroidMeter for searching and measuring embedded Web pages.}\label{fig:tool}
\end{figure}

We address these two challenges by building an automation tool DroidMeter
to analyze the performance of embedded pages. Figure~\ref{fig:tool} shows the architecture of DroidMeter, which consists of three components: {\em explorer, replayer}, and {\em page collector}.

Given an app under analysis, the explorer launches the app and traverse to app pages. During the traversal, the explorer builds a state machine of page transition. If the tool lands on an activity that embeds a Web page (for example, pages P4 and P5 shown in dashed line), the tool generates a replay script. This replay script records the set of steps from the beginning of the app to the activity with the embedded Web page. Using the script, the replayer can trigger the embedded Web page. Meanwhile, the page collector retrieves performance metrics about the embedded Web pages. We give the details of each component as follows.



\subsubsection{Explorer}
The explorer is a modified version of the Monkey tool~\cite{Monkey} provided by the Android SDK. The Monkey tool is an automation tool that can traverse through the different activities of the app but randomly clicking and typing on the UI of the app. Similar to the Monkey tool, the explorer launches the app and performs a depth-first search to traverse the app activities. Searching essentially involves clicking on various UI elements to either trigger a change in UI or trigger a navigation to a different activity. The explorer performs dynamic analysis to identify embedded Web pages in an activity. Specifically the explorer parses the UI structure of each activity to identify if the activity contains an embedded Web page.

There are two key problems with existing automation tools such as the Monkey tool~\cite{Monkey}. The first is coverage and the second is repeated traversal to the same activity. The DroidMeter explorer avoid both by (a) maintaining a state machine and (b) using the UI structure of each activity to distinguish between the activities. DroidMeter build a page transition model. The model is a state machine where each node represents an app page and the edge represents the page transition via events.

DroidMeter is able to differentiate between the activities using the structure of the UI tree~\cite{uitree}. The UI tree consists of UI elements of an app page similar to a DOM tree of Web pages. Since the structure does not change significantly during replay, DroidMeter is able to faithfully reach the target embedded Web pages.

\subsubsection{Replayer}
At the end of the exploration, DroidMeter generates a script to reach all activities with an embedded Web page. For example, in Figure~\ref{fig:tool} page P4 is an app page with embedded Web page, and the replay script of P4 is to first trigger event e1 on page P1 and then trigger event e3 on page P2. To open the page, the replayer component triggers each event according to the script. The state machine we built as part of the explorer is used to locate the UI elements that will trigger each transformation.

\subsubsection{Page Collector}
During the replay, the page collector retrieves information about the loading procedure of both the embedded Web page and the Android native activity.
To retrieve information of embedded Web pages, we port the chrome debugging tools, specifically to expose debug APIs. The debugging port on the mobile device is forwarded to a local port from which we  obtain information about resources downloaded, timing information, and information about HTTP request/response.

\subsection{Measuring Metrics}

In order to compare the performance of mobile apps with and without embedded Web pages, DroidMeter measures the page load time of both embedded Web pages and native apps. The page collector ports the chrome debugging tools that provides information about the page load time.

DroidMeter uses the APIs in the network domain of Chrome remote debugging protocol~\cite{networkdomain}. Network domain provides APIs to track network activities of the Web page. It exposes information about http, file, data and other requests and responses, their headers, bodies, timing, etc.
We organize the retrieved information based on the HAR (HTTP ARchive) specification~\cite{har} for analysis.

To measure the equivalent of the page load time of mobile apps, we need to measure the time that elapses between the beginning of the activity and the time when the callback method onCreate() finishes executing. We use the Xposed Framework to inject some code before and after the  onCreate() method to perform this measurement. To get the timestamp of when the application activity launches, we use the method \texttt{reportFullyDrawn()}. This method is designed for debugging and optimizing purpose and does not affect the behavior of the activity. But in each activity, this method can be called only once. Repeated calls to this method will lead to all the calls except the first one being ignored. We inject this method call into activities just before the activity's onCreate() method is called, alongside with codes which will print the current timestamp into logcat. With these two timing data, we can calculate the timestamp of the very beginning of an activity, then we can know the activity's timing information.

To measure the SpeedIndex~\cite{speedindex} of loadings of both embedded web pages and native app pages, DroidMeter just records the screen and calculates the SpeedIndex from the recorded video. DroidMeter uses the shell command \texttt{screenrecord} to record the screen, and uses the utility \texttt{visualmetrics}~\cite{visualmetrics} to calculate the SpeedIndex value.

The flexible architecture of DroidMeter makes it easy for users to extend it to support new measurement or new metrics.

\section{Case Study}\label{sec:casestudy}

In this section, we study a real-life execution of DroidMeter over an Android App, to illustrate the usage and bahavior of DroidMeter.

\begin{figure}
\centering
\minipage{0.49\textwidth}
  \includegraphics[width=\linewidth]{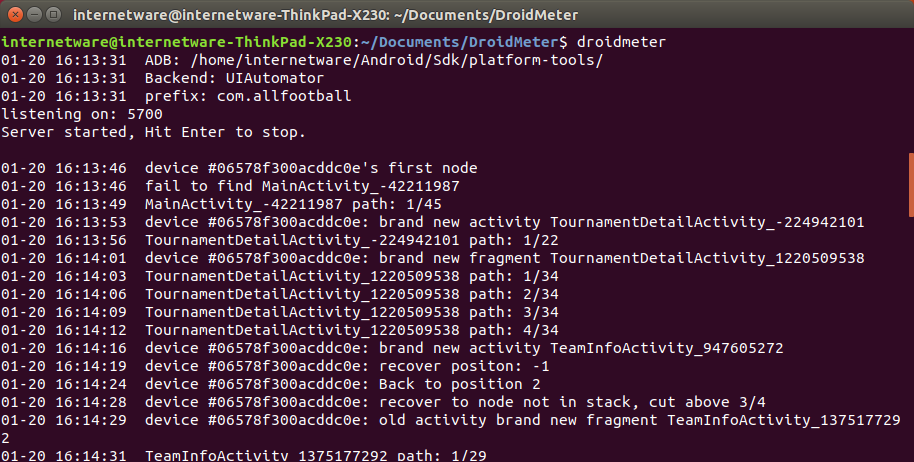}
  \caption{Exploring}~\label{fig:casestudy:exploring}
\endminipage\hfill
\minipage{0.49\textwidth}
  \includegraphics[width=\linewidth]{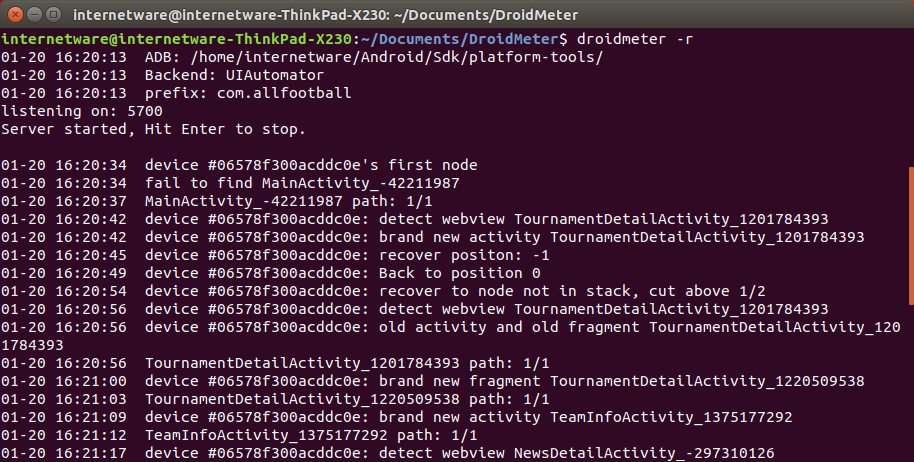}
  \caption{Replaying}~\label{fig:casestudy:replaying}
\endminipage\hfill
\end{figure}

\begin{figure}
  \centering
  \includegraphics[width=0.5\textwidth]{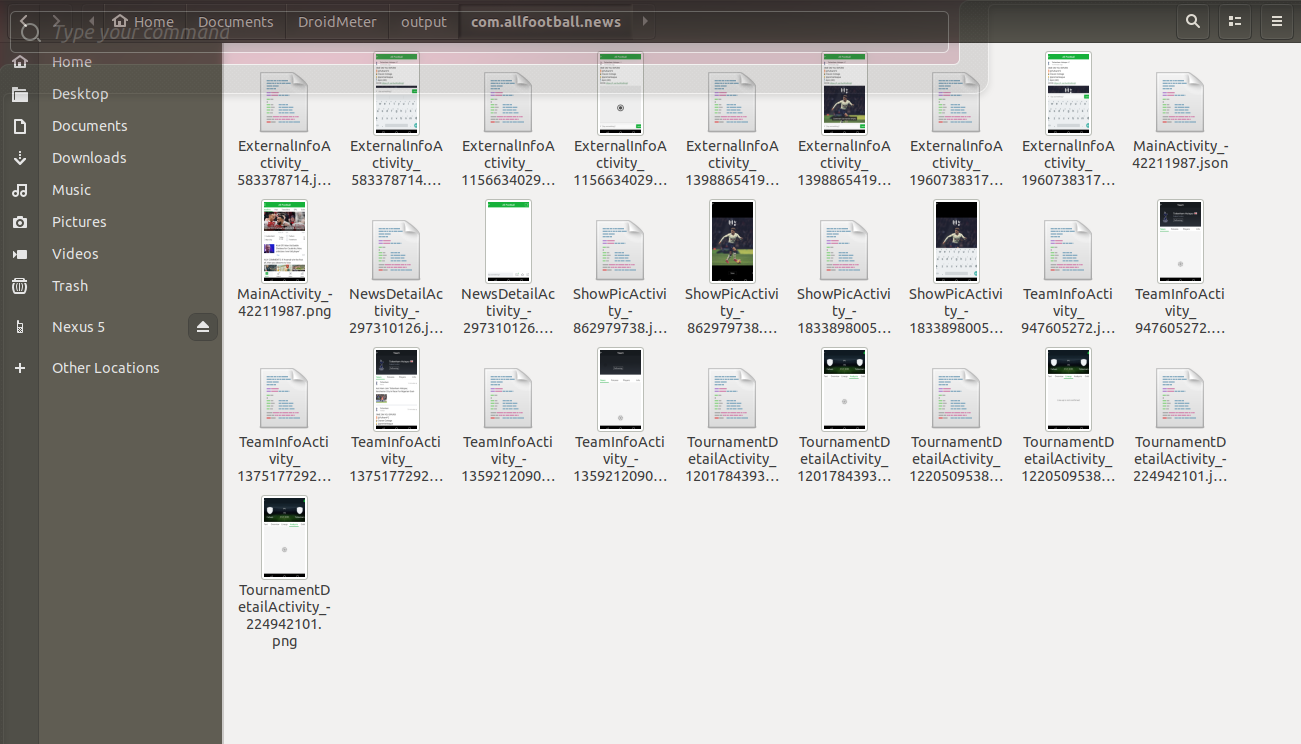}
  \caption{DroidMeter output}~\label{fig:casestudy:output}
\end{figure}

\subsection{Searching for Embedded Web Pages}

We use the app {\em All Football} as the app being studied.
First, we set up the running environment of DroidMeter. 
Before experiment we should connect a mobile phone and a computer by a USB wire and install the target app in the mobile phone.
To configure the testing target for DroidMeter, we specify the mobile device and mobile app in a configuration file. We leave some other trivial options as their default values.
After we have well prepared the input for DroidMeter, we execute DroidMeter in one shell command.

When DroidMeter starts, it reads the configuration file to get the basic information.
Then, DroidMeter controls the mobile device to start the app to be tested.
Once the app is started, DroidMeter browses the app and checks the layout of the on-screen activity. 
If DroidMeter finds a WebView widget in the layout of current activity, it records some basic information of the WebView and the activity, and generate a replay script which directly leads to this activity.
When DroidMeter meets the same activity twice, it will recognize the identity of the activity and avoid conducting redundant tests.
When the maximum exploration time reached, DroidMeter stops the exploration, saves the exploration result into a specified folder and exits.
After it exits, user can manually check the output folder and analyze the exploration results. These results are also needed when DroidMeter replays operations and collects metrics.

Figure~\ref{fig:casestudy:exploring} shows the command line output of DroidMeter when it is exploring the app {\em All Football}. When started, DroidMeter reads the configuration file, obtains the entry activity of the app {\em All Football}, then it browses the app and records the activities that it meets.
Meanwhile, DroidMeter listens to a specific TCP port, so a user can send command to DroidMeter. A user can let DroidMeter save the result by sending a ``save'' command, and stop DroidMeter by pressing the ``Ctrl-C'' key.

\subsection{Replaying and Measurement}
After DroidMeter explores an app and saves the replay scripts, DroidMeter can replay the operation and navigate to some specific activity in the app again according to the replay scripts.
After a user triggers DroidMeter and orders it to replay, DroidMeter reads the replay scripts, constructs an ``operation sequence'' which leads to an activity that contains a WebView. Then DroidMeter perform the operation sequence to reach the target activity, or the target WebView. During the replaying, DroidMeter checks whether the app state is correct. If the app state is inconsistent with the recorded state, DroidMeter rewinds the last operation and retries it. Finally, DroidMeter will reach the target activity, or DroidMeter will report that the target activity is unreachable.

DroidMeter measures the user-specified metrics while DroidMeter replays the operations. During replaying, before each operation, DroidMeter prepares to measure; after each operation, DroidMeter start to measure; then after a while DroidMeter save the metrics into a folder.

Figure~\ref{fig:casestudy:replaying} shows the command line output of DroidMeter when it is replaying script which leads to the recorded WebView in the app {\em All Football}. DroidMeter replays the operation saved in the scripts and prints some information in command line.

Figure~\ref{fig:casestudy:output} shows the output of DroidMeter. Currently, for each activity, DroidMeter records the screenshot of the activity and some other basic information.
User can easily extend DroidMeter to support other metrics.

\section{Related Work}\label{sec:related}
In this section, we survey the related research efforts on characterizing Web browsing performance and studying Android WebView.

\subsection{Web Browsing Measurement Studies}
Understanding the characteristic and performance of Web browsing is essential for providing better user experience. For desktop browsing, there has been considerable effort on studying Web traffic~\cite{Ihm:IMC2011,Ager:IMC2011} over the Internet and the characterizing Web page resources~\cite{Butkiewicz:IMC2011}. Wang et al.~\cite{Wang:NSDI2013} designed WProf to analyze the dependency of the activities in a page load and identify the critical path. 

In terms of mobile browsers, measurement studies focus especially on resource utilization~\cite{Qian:Mobisys2014}, caching mechanism~\cite{Qian:Mobisys2012,Ma:WWW2015}, and energy consumption~\cite{Thiagarajan:WWW2012,Bui:MobiCom2015}. Nejati et al.~\cite{Nejati:WWW2016} designed WProf-M to analyze the key bottlenecks in mobile browsers. 

There are a lot of performance metrics which reflect page loading performance in many aspects. Bocchi et al.~\cite{Bocchi:2016} summarize many useful metrics reveal the relationships among them. Netravali et al.~\cite{Netravali:NSDI2018} propose a new metric called Ready Index, which captures the load time of a Web page with respect to its interactivity.

\subsection{Web Browsing Optimization Studies}
There are many works concentrating on improving the performance of Web Browsing.

Liu et al.~\cite{DBLP:journals/tmc/LiuMDLXH17} propose a method that leverages resource-packaging to reduce the redundant transfers in mobile Web browsing. They ~\cite{DBLP:journals/tmc/LiuMWLXH17} also optimize the resource loading procedure in mobile Web browsing.
Ma et al.~\cite{Ma:WWW2015} measures and analyzes the performance of Web cache in mobile browsing. Surprisingly, they find that the utilization of Web cache in Web browsing is fairly low.
Further, Liu et al.~\cite{liu_demystifying_2016} study the reason of the bad performance of Web cache. 

%

\section{Conclusion}\label{sec:conclusion}

The goal of this work is to characterize the performance and resource usage of embedded Web pages.
Embedded Web pages are pages that are embedded within mobile apps. Although the security of embedded Web pages are well-studied, little is known about their performance characteristics.
We have built tools to study the characteristics of embedded Web pages at scale.


\bibliography{droidmeter}

\begin{thebibliography}{10}

\bibitem{ChromeRemote}
Chrome remote debugging.
\newblock \url{https://developer.chrome.com/devtools/docs/remote-debugging},
  2018.

\bibitem{uitree}
Hierarchy viewer.
\newblock \url{https://developer.android.com/studio/profile/hierarchy-viewer},
  2018.

\bibitem{har}
Http archive (har) format.
\newblock
  \url{https://dvcs.w3.org/hg/webperf/raw-file/tip/specs/HAR/Overview.html},
  2018.

\bibitem{networkdomain}
Network domain in chrome remote debugging protocol.
\newblock \url{https://chromedevtools.github.io/devtools-protocol/tot/Network},
  2018.

\bibitem{speedindex}
Speed index.
\newblock
  \url{https://sites.google.com/a/webpagetest.org/docs/using-webpagetest/metrics/speed-index},
  2018.

\bibitem{Monkey}
Ui/application exerciser monkey.
\newblock \url{https://developer.android.com/studio/test/monkey}, 2018.

\bibitem{visualmetrics}
visualmetrics.
\newblock \url{https://github.com/WPO-Foundation/visualmetrics}, 2018.

\bibitem{webkit}
Wandoujia app store.
\newblock \url{https://webkit.org/}, 2018.

\bibitem{chromium}
Wandoujia app store.
\newblock \url{https://www.chromium.org/}, 2018.

\bibitem{WebView}
Webview on android.
\newblock \url{https://developer.android.com/reference/android/webkit/WebView},
  2018.

\bibitem{Ager:IMC2011}
B.~Ager, W.~M{\"{u}}hlbauer, G.~Smaragdakis, and S.~Uhlig.
\newblock Web content cartography.
\newblock In {\em Proceedings of the 11th {ACM} {SIGCOMM} Internet Measurement
  Conference, {IMC} 2011}, pages 585--600, 2011.

\bibitem{Bocchi:2016}
E.~Bocchi, L.~D. Cicco, and D.~Rossi.
\newblock Measuring the quality of experience of web users.
\newblock In {\em Proceedings of the 2016 Workshop on QoE-based Analysis and
  Management of Data Communication Networks}, pages 37--42, 2016.

\bibitem{Bui:MobiCom2015}
D.~H. Bui, Y.~Liu, H.~Kim, I.~Shin, and F.~Zhao.
\newblock Rethinking energy-performance trade-off in mobile web page loading.
\newblock In {\em Proceedings of the 21st Annual International Conference on
  Mobile Computing and Networking, MobiCom 2015}, pages 14--26, 2015.

\bibitem{Butkiewicz:IMC2011}
M.~Butkiewicz, H.~V. Madhyastha, and V.~Sekar.
\newblock Understanding website complexity: measurements, metrics, and
  implications.
\newblock In {\em Proceedings of the 11th {ACM} {SIGCOMM} Internet Measurement
  Conference, {IMC} 2011}, pages 313--328, 2011.

\bibitem{Ihm:IMC2011}
S.~Ihm and V.~S. Pai.
\newblock Towards understanding modern web traffic.
\newblock In {\em Proceedings of the 11th {ACM} {SIGCOMM} Internet Measurement
  Conference, {IMC} 2011}, pages 295--312, 2011.

\bibitem{DBLP:journals/tmc/LiuMDLXH17}
X.~Liu, Y.~Ma, S.~Dong, Y.~Liu, T.~Xie, and G.~Huang.
\newblock Rewap: Reducing redundant transfers for mobile web browsing via
  app-specific resource packaging.
\newblock {\em {IEEE} Transactions on Mobile Computing}, 16(9):2625--2638,
  2017.

\bibitem{liu_demystifying_2016}
X.~Liu, Y.~Ma, Y.~Liu, T.~Xie, and G.~Huang.
\newblock Demystifying the {Imperfect} {Client}-{Side} {Cache} {Performance} of
  {Mobile} {Web} {Browsing}.
\newblock {\em IEEE Transactions on Mobile Computing}, 15(9):2206--2220, Sept.
  2016.

\bibitem{DBLP:journals/tmc/LiuMWLXH17}
X.~Liu, Y.~Ma, X.~Wang, Y.~Liu, T.~Xie, and G.~Huang.
\newblock Swarovsky: Optimizing resource loading for mobile web browsing.
\newblock {\em {IEEE} Transactions on Mobile Computing}, 16(10):2941--2954,
  2017.

\bibitem{Ma:TMC2018}
Y.~Ma, X.~Liu, Y.~Liu, Y.~Liu, and G.~Huang.
\newblock A tale of two fashions: An empirical study on the performance of
  native apps and web apps on android.
\newblock {\em {IEEE} Transactions on Mobile Computing}, 17(5):990--1003, 2018.

\bibitem{Ma:WWW2015}
Y.~Ma, X.~Liu, S.~Zhang, R.~Xiang, Y.~Liu, and T.~Xie.
\newblock Measurement and analysis of mobile web cache performance.
\newblock In {\em Proceedings of the 24th International Conference on World
  Wide Web, {WWW} 2015}, pages 691--701, 2015.

\bibitem{Nejati:WWW2016}
J.~Nejati and A.~Balasubramanian.
\newblock An in-depth study of mobile browser performance.
\newblock In {\em Proceedings of the 25th International Conference on World
  Wide Web, {WWW} 2016}, pages 1305--1315, 2016.

\bibitem{Netravali:NSDI2018}
R.~Netravali, V.~Nathan, J.~Mickens, and H.~Balakrishnan.
\newblock Vesper: Measuring time-to-interactivity for web pages.
\newblock In {\em Proceedings of the 15th {USENIX} Symposium on Networked
  Systems Design and Implementation {NSDI} 2018}, pages 217--231, 2018.

\bibitem{Qian:Mobisys2012}
F.~Qian, K.~S. Quah, J.~Huang, J.~Erman, A.~Gerber, Z.~M. Mao, S.~Sen, and
  O.~Spatscheck.
\newblock Web caching on smartphones: ideal vs. reality.
\newblock In {\em Proceedings of the 10th International Conference on Mobile
  Systems, Applications, and Services, MobiSys 2012}, pages 127--140, 2012.

\bibitem{Qian:Mobisys2014}
F.~Qian, S.~Sen, and O.~Spatscheck.
\newblock Characterizing resource usage for mobile web browsing.
\newblock In {\em Proceedings of the 12th Annual International Conference on
  Mobile Systems, Applications, and Services, MobiSys 2014}, pages 218--231,
  2014.

\bibitem{Serrano:Software2013}
N.~Serrano, J.~Hernantes, and G.~Gallardo.
\newblock Mobile web apps.
\newblock {\em {IEEE} Software}, 30(5):22--27, 2013.

\bibitem{Thiagarajan:WWW2012}
N.~Thiagarajan, G.~Aggarwal, A.~Nicoara, D.~Boneh, and J.~P. Singh.
\newblock Who killed my battery?: analyzing mobile browser energy consumption.
\newblock In {\em Proceedings of the 21st World Wide Web Conference 2012, {WWW}
  2012}, pages 41--50, 2012.

\bibitem{Wang:NSDI2013}
X.~S. Wang, A.~Balasubramanian, A.~Krishnamurthy, and D.~Wetherall.
\newblock Demystifying page load performance with wprof.
\newblock In {\em Proceedings of the 10th {USENIX} Symposium on Networked
  Systems Design and Implementation, {NSDI} 2013}, pages 473--485, 2013.

\end{thebibliography}

\bibliographystyle{abbrv}

\end{document}